# Does Campaigning on Social Media Make a Difference? Evidence from candidate use of Twitter during the 2015 and 2017 UK Elections[1]

*Jonathan Bright, Scott Hale, Bharath Ganesh, Andrew Bulovsky, Helen Margetts, Phil Howard.*

## Abstract

Social media are now a routine part of political campaigns all over the world. However, studies of the impact of campaigning on social platform have thus far been limited to cross-section datasets from one election period which are vulnerable to unobserved variable bias. Hence empirical evidence on the effectiveness of political social media activity is thin. We address this deficit by analysing a novel panel dataset of political Twitter activity in the 2015 and 2017 elections in the United Kingdom. We find that Twitter based campaigning does seem to help win votes, a finding which is consistent across a variety of different model specifications including a first difference regression. The impact of Twitter use is small in absolute terms, though comparable with that of campaign spending. Our data also support the idea that effects are mediated through other communication channels, hence challenging the relevance of engaging in an interactive fashion.

[1] Many people provided invaluable comments and suggestions on earlier drafts of this paper. We would like to thank in particular Rebecca Eynon, Leticia Bode, and participants at the Reuters-OII Seminar of June 2017.

## 1 Introduction

Over the last decade or so, campaigning on social media platforms (whereby political candidates create and maintain profiles on social media networks during electoral campaign periods) has become a core feature of contemporary political systems all around the world (Dimitrova and Matthes 2018). Initially a niche pursuit, now parties of all sizes and political leanings are making heavy use of these tools to promote their messages and candidates (for some examples see Bode & Epstein, 2015; Tan, Tng, & Yeo, 2016; Grusell & Nord, 2016; Quinlan, Gummer, Roßmann, & Wolf, 2017; Samuel-Azran, Yarchi, & Wolsfeld, 2015; Koc-Michalska et al., 2016; Lilleker et al., 2011; Ramos-Serrano, Gomez & Pineda 2016). Indeed, this use of social media is now likely to be incorporated as one political communication tool among many others in professionally-run campaigns (Stromer-Galley, 2014, Štětka, Lilleker, Tenscher, & Jalali, 2014; Quinlan et al., 2017), which also includes a variety of other interactive web technologies (see Koc-Michalska et al., 2016, p. 344-5; Kruikemeier, 2014, p. 135; Van Noort, Vliegenhart, & Kruikemeier 2016, p. 360).

This article addresses the simple but foundational question of whether the effort put into social media platforms is effective in terms of winning votes. There is a small amount of literature that has already sought to treat this subject, which has produced diverging results. Some researchers have argued that there is a positive relationship between social media use and vote outcomes. For example, Kruikemeier (2014) found that politicians using Twitter in the Dutch national elections of 2010 were more likely to be successful; Bode and Epstein (2015) look at aggregate measures of online influence ("Klout" scores) in the context of the US 2012 elections, also finding that politicians with more Klout were more successful; LaMarre and Suzuki-Lambrecht (2013) argued that 'candidates' Twitter use significantly increased their odds of winning' in the 2010 US House of Representatives elections; Vergeer, Hermans and Sams (2011) also claimed that 'there is indeed a real effect of micro–blogging activity on the number of votes candidates receive' in the context of the 2009 European Parliament elections; and most recently Bene (2018) found a correlation between shares on candidate Facebook pages and vote outcomes in the 2014 Hungarian elections.

Others have argued that there is no relationship: for example, Vaccari and Nielsen (2013) produced a study showing that online popularity on Twitter does not correlate with vote share (though they did find a relationship in the case of Facebook), whilst Baxter and Marcella also concluded that there was little evidence of a causal relationship between social media use and vote share outcomes in the 2011 Scottish Parliamentary election (Baxter & Marcella, 2013, p. 203). Also notable in this regard is the fact that many systematic efforts to predict elections by studying social media activity have ended in failure (see, for example, McGregor, Morau & Molyneux, 2017, p. 10; Murthy, 2015; Gayo-Avello, 2013; Burnap, Gibson, Sloan, Southern & Williams, 2016; Jungherr, Schoen, Posegga & Jürgens, 2016).

One potential reason for these diverging findings is that, to our knowledge, all of the studies on the issue so far are cross-sectional, in that they study one election in one country. Cross-sectional studies of the impact of social media (or indeed any campaigning technique) on vote share outcomes are potentially problematic because political campaigns will adopt multiple strategies in order to generate votes. Any measure of social media use may hence also act as a proxy measure for other types of engagement activities, or the overall professionalization of the campaign, or the campaign skill and charisma of the candidate (Levitt, 1994). Models of the relationship between social media use and vote outcomes will therefore be vulnerable to potential unobserved variable bias: social media use may simply be heavier amongst 'better' campaigns. Theoretically relevant control variables such as level of campaign spending can ameliorate the situation but not remove the problem entirely. For example, ephemeral characteristics such as the skill of the candidate or the dedication of campaign staff will be very hard to capture in this fashion.

This study hence seeks to build on and extend this previous work, by offering a stronger test of the impact of social media use on vote share outcomes. We focus particularly on Twitter, which is one of the most widely adopted social media platforms for political purposes (Jungherr, 2016), with high penetration rates in political systems all over the world: published research has documented its use in electoral competitions in countries as diverse as Pakistan (Ahmed & Skoric, 2014), Sweden (Larsson & Moe, 2012), New Zealand (O'Neill, 2012), Indonesia (Amirullah, Komp & Nurhadryani, 2013), Spain (Aragón et al. 2013), Italy (Bentivegna, 2014), Australia (Bruns & Burgess 2011), Norway (Enli & Skogerbø, 2013) and South Korea (Hsu & Park, 2012), to take but a few examples.

Our study leverages the fact that the United Kingdom [UK] has recently held two electoral competitions in relatively quick succession (in 2015 and 2017) to create a novel pseudo-panel dataset of the Twitter use of almost 6,000 politicians during two separate electoral periods. We observe the extent to which the level of Twitter activity in both periods correlates with vote outcomes (in both cross-sectional and pooled time-series regressions), and test whether this effect is robust to controls for potential confounding variables. We also estimate a series of first difference models on the limited subset of around 800 politicians who competed in both elections, a design which allows us to eliminate a wide range of unobserved variables such as the candidate's campaigning skill or whether they are located in a safe seat. Our results demonstrate that campaigning on Twitter does indeed 'make a difference'. We also show that larger parties are apparently more able to take advantage of the platform, and we undermine existing literature which has claimed that engaging with the platform in an 'interactive' style (by engaging in conversations and replying to citizens) is the most effective method of social media campaigning. We also provide suggestive evidence that the impact of Twitter is mediated through other channels (such as the mainstream media).

The remainder of the article is structured in the following way. Section 2 addresses theory about the political use of social media in general and outlines why campaigning on social media platforms might make a difference. It also specifies interaction effects that ought to moderate the extent to which social media campaigning is useful, and thus elaborates some hypotheses to be tested. Section 3 describes our methods, data collection and variable operationalisation. Finally, Section 4 presents our analysis and findings from the study.

## 2 Political Campaigning on Twitter: Mechanisms and moderators of influence

As a variety of authors have noted, social media platforms such as Twitter have altered the 'media ecology' surrounding the public during political campaigns (e.g. Bimber, 2014; Gurevtich, Coleman, & Blumler, 2009), and these changes have opened up a variety of mechanisms through which a political candidate might influence voting outcomes. We split these into two basic types of pathway: *direct* and *indirect mobilisation*

In terms of *direct mobilisation*, it is possible that the use of Twitter might open a channel of direct communication between the political candidate in question and citizens who are eligible to vote: individuals using the platform might see messages posted by candidates, either because they are directly connected to the candidate themselves (something which is increasingly common – see Pew Research Center, 2014), or because messages posted by the candidate are shared by their social circle. When voters see such messages, they are informed about both the existence of an upcoming election and that the candidate is someone they could vote for (some evidence for this is provided by Gottfried, Hardy, Holber, Winneg, & Jamieson, 2017, who have shown that knowledge of political campaigns correlates with social media use). Voters may be motivated by direct appeals to participate, something which Vaccari (2017) has shown to be impactful (though direct political appeals across social media can be treated with scepticism - see Boerman & Kruikemeier, 2016). They may also be informed about characteristics of the candidate that might help sway their vote, for example they will receive 'social information' about the number of friends or followers a given candidate has (Margetts, John, Hale, & Yasseri, 2015). Research on Facebook has shown this to be politically powerful (Bond et al., 2012). There is also evidence that campaigning through social media platforms such as Twitter leads to voters feeling more direct connections with politicians (Lee & Shin, 2012), something which is also likely to boost their willingness to vote.

In addition to these direct connections, there is also the possibility of *indirect mobilisation*. As Murthy notes (2015), campaigning on platforms such as Twitter can be used to generate traditional press coverage, as the press themselves are covering social media, through a process sometimes described as inter-media agenda setting (see also Conway, Kenski, & Wang, 2015; Anstead & O'Loughlin, 2015). This press coverage may then stimulate further interest on the part of voters (Yasseri & Bright, 2016), whilst news articles may be shared via social media by either the candidate themselves or their supporters (Bright, 2016). Twitter messages may also be heard by the 'politically engaged public' (Weaver et al., 2018), who may well be significant for mobilising further swathes of opinion. For example, they could act as 'opinion leaders': small groups of highly politically active people who then in turn go on to influence others within their own social circle (Dubois & Gaffney, 2014; Borge Bravo & Esteve del Valle, 2017; Bermudez, Bright, Pilet & Soubiran, 2016). Twitter use could also be a way of connecting to political activists, who might themselves contribute to further campaign efforts (see Bastos & Mercea, 2015). zThe possibility for both direct and indirect mobilisation effects through Twitter lead us to develop the first, basic hypothesis which is to be tested in the article:

*H1: Candidates who make more use of Twitter during an election period will win more votes*

In addition to these influence mechanisms, there are good reasons to expect that the possibility of realising electoral gains from Twitter use is likely to be moderated by other factors. One example here concerns the size of a candidate's potential audience on Twitter (this is, clearly, of more relevance to the *direct* mobilisation mechanism than it is to the *indirect* one). This may relate to the extent to which Twitter is actually used amongst their potential voters. Many elections, such as the ones we investigate here, are fought in geographically constrained areas called constituencies, which may well have demographic profiles that vary considerably from the country at large. And demographics are known to be drivers of differential use of Twitter: in the UK, users of the platform are known to be considerably younger than average, considerably more likely to have higher levels of educational qualification and also more likely to be economically better off (findings from Blank, 2017, pp. 683-686; Malik, Lamba, Nakos & Pfeffer, 2015 also reach similar conclusions). Research has also suggested that they may be more likely to live in urban areas (Hecht & Stephens, 2014; Barberá & Rivero, 2014). These different demographic profiles of Twitter users mean that we might expect the penetration of Twitter to vary considerably from constituency to constituency, and of course we would expect the impact of Twitter campaigning to be greater in areas with higher levels of Twitter use.

The potential size of a politician's audience also relates to the number of 'followers' they have on the platform (people who have chosen to receive output from the candidate). Followers are the people who are most likely to see the tweets produced by the campaign: hence having more of them indicates greater potential for the direct mobilisation mechanism described above (and research has found that increased follower levels for a candidate seem to make a difference – see Vergeer, Hermans & Sams, 2011). Of course, not every follower sees every tweet (Twitter's own analytics suggested that tweeting two to three times a day would allow a typical account to reach 30% of its followers every week – see Twitter, 2014). Nevertheless, more followers ought to indicate increased influence. Larger follower counts will also send a greater signal of political viability (which is consequential for both *direct* and *indirect* mobilisation). In this respect, it is interesting to note that many politicians operate what has been described as a 'campaign and withdraw' strategy on social media (Enli & Skogerbø, 2013), whereby accounts are set up shortly before the election takes place and often deleted shortly afterwards. This is clearly less favourable to the development of large follower bases. These theories on audience size lead us to the second hypothesis to be tested:

*H2: Politicians with larger audiences will benefit more from campaigning on Twitter*

In addition to audience size, a further line of research suggests that there may be party level differences in the effectiveness of Twitter use. One of the earliest debates within the digital politics literature concerned that between the theses of 'equalisation' and 'normalisation' (see e.g. Gibson & McAllister, 2015). The equalisation thesis suggests that minor parties are more likely to make use of digital technologies such as Twitter, because the low cost of communication on the platforms enables them to level the playing field when compared with more established actors who dominate traditional media channels (Enli & Skogerbø, 2013, p.759; Koc-Michalska, Lilleker, Smith & Weissmann, 2016; Larsson & Moe, 2014). The normalisation thesis, by contrast, suggests that large parties will make use of their existing resource advantages to also dominate these new channels.

While these debates largely relate to who is using the technology, Gibson and McAllister (2015) also specifically relate them to who *benefits* from the technology, providing evidence that green parties appeared to benefit more from using social media than their more mainstream competitors in Australian elections, a finding they relate to the fact that these voters are typically drawn from younger demographics (see Gibson & McAllister, 2015, p.542). Recent studies showing that social media have a positive impact on the youth vote seem to support this idea (Vaccari, 2017; Garzia, Trechsel, & De Angelis 2017; Aldrich, Gibson, Cantijoch & Konitzer, 2015).[2] Hence, we develop our next hypothesis:

*H3: Politicians from smaller parties will benefit more from campaigning on Twitter*

Another potentially relevant factor concerns the political makeup of those interacting with the politician. A recurring line of research in studies of digital politics concerns what could be referred to as the 'echo chamber' thesis (see e.g. Conover et al., 2012; Bright, 2018): the idea that on social media, conversations are largely held amongst those who already agree with each other. In the case of Twitter in the UK in particular, Weaver et al. (2018) have argued that there is a 'dominant state of partisan segregation' on the platform. It could be that politicians who have more heterogeneous political networks are better able to reach and influence people who were not already considering voting for them (though of course mobilising those who already agree with you may have important benefits in itself by increasing the likelihood they will turn out to vote and by perhaps convincing them to engage actively in a campaign). This leads us to our fourth hypothesis:

*H4: Politicians with more heterogeneous political networks will benefit more from campaigning on Twitter*

A final aspect to consider is the style of engagement that politicians make with Twitter, a subject that has generated a considerable literature (see e.g. Graham, Jackson & Broersma, 2014; Evans, Cordova & Sipole, 2014; Golbeck, Grimes & Rogers, 2010; Jackson & Lilleker, 2011; Bulovsky, 2018). Graham et al. (2014, see especially pp. 697-698) provide a useful general breakdown of issues that matter here: a political tweet can be classified in terms of whether it is simply a 'broadcast' message reporting on a politician's activities or viewpoints or whether it involves some sort of 'interaction' (e.g. is a reply to another user or makes some kind of attempt to engage in a discourse). Empirical research has consistently critiqued politicians for engaging with Twitter in a more broadcast style, and make less use of its interactive features (see Jungherr, 2017, p. 76; Ross & Bürger, 2014), with larger political forces apparently even less likely to interact (Heiss, Schmuck & Matthes, 2018), thus missing out on the potential of the platform. 'Populist' parties are also, apparently, less likely to engage in interaction (Jacobs & Spierings, 2018). When there is some kind of interactivity and engagement, it typically takes place with other politicians, especially members of the same party (see Livne et al. 2011). However, the finding is not necessarily consistent for all parties: Heiss et al. (2018) have argued that politicians from smaller parties were more likely to directly interact with and respond to users. This seems important because the potential influence of the *direct mobilisation* mechanism as described above might depend at least partly on the extent to which politicians are actually seen as responsive to individual voters, and indeed literature on the subject has suggested that using a more interactive style offers a 'vote dividend' (Koc-Michalska, Lilleker & Smith, 2016). This leads us to our final hypothesis:

*H5: Politicians with more interactive tweeting strategies will benefit more from campaigning on Twitter*

## 3 Methods

Our study is based on data from the 2015 and 2017 UK national elections. In these elections, political parties compete by fielding candidates in 649 individual electoral districts, known as "constituencies", which each contain approximately 70,000 people who are registered to vote[3]. In our dataset, we include all candidates who stood for any party that managed to win at least one seat in at least one of the two elections we observed[4]. We did not include the 18 Northern Irish constituencies, where parties from the rest of the UK do not typically campaign and where electoral politics has a local focus that is mostly independent from national political dynamics, hence our geographic focus is limited to 631 constituencies in England, Scotland and Wales. In total, our dataset contains almost 6,000 candidate observations campaigning for one of seven different parties in one of these constituencies in at least one of the two elections. The dependent variable for our study is the percentage of overall votes gained by each candidate in the constituency in which they were competing. The main method employed in this paper, consistent with a variety of other studies of the campaign effects of digital technologies (Kruikemeier, 2014; Koc-Michalska et al., 2016; Van Noort et al., 2016), is to observe the extent to which effort put into Twitter by candidates in a political campaign correlates with these vote outcomes.

A subset of these politicians campaigned in both elections, something which allows us to treat the data as a pseudo-panel dataset. This means that, in addition to running cross sectional and pooled models, we also run first difference models observing change in outcomes within individual candidates (Wooldridge, 2012, p. 279). This approach has been used previously in studies of campaign effects as a way of eliminating (time-invariant) unobserved variables such as charisma which are of obvious importance and yet very difficult to measure (see e.g. Levitt, 1994). Of course, it is worth noting that the subset of politicians who campaigned in both elections is biased towards those who are more successful: 52% of those who returned in 2017 had won their seat in 2015, whilst the rest typically scored much higher than the average candidate in the 2015 election. However, the fact that we can run models on two waves of cross sectional data and take a panel approach means that overall we are positioned to offer a strong test of the potential impact of Twitter on voting outcomes.

Our main independent variables are based on patterns of candidate activity on Twitter. This is a highly-used social media platform, which had approximately 14.8 million UK based users in 2015, a number which rose to 16.4 million in 2017 (Statista, 2017). This is relatively large considering the UK had a population of around 65 million in this time period, of which around 45 million were eligible to vote (The Electoral Commission 2016). Users on the platform can distribute short messages, known as "tweets", to individuals who have chosen to follow them. They can also "reply" to messages created by others, and "retweet" messages created by others (which involves rebroadcasting someone else's tweet to your own followers). The distinguishing feature of the platform is that it is highly public, meaning that anyone can choose to follow anyone else. This focus on only one social media platform does create a limitation of course, as we do not know how Twitter activity may correlate with activity on other social platforms. We will return to this point in the discussion.

---

[3] There are actually 650 constituencies in total, but one is dedicated to the ceremonial position of the "speaker of the House", a politician who organises the business of parliament and hence who traditionally runs unopposed. The exact number of voters varies slightly by constituency. For details, see: https://www.ons.gov.uk/peoplepopulationandcommunity/elections/electoralregistration/bulletins/electoralstatisticsforuk/2014-05-01

[4] These parties were: the Conservative Party, the Labour Party, the Scottish National Party, the Liberal Democrats, Plaid Cymru, the Green Party and the UK Independence Party.

We collected a number of variables from Twitter. We will describe each of these in turn here, according to the hypothesis they are designed to test.

## H1: Twitter Usage

To address H1, we collected basic metrics about the candidate's usage of Twitter. In order to collect data on candidate Twitter activity, we made use of data from the civic technology organisation *DemocracyClub*[5], which released information on the Twitter usernames of UK political candidates. We conducted an independent verification of the data, which overall was found to be 93% accurate.[6] Using the list of candidates provided by this organisation, we first recorded whether political candidates in our dataset had a Twitter account or not. For those that did, we tracked activity on Twitter[7] from the day after official candidate registration closed up to and including the day before each election, a period of 24 days in total.[8] This allowed us to record the number of contributions candidates made to the platform during the campaign period (including replies, retweets and 'original' tweets.

## H2: Audience Size

We address the audience size hypothesis in two ways. First, we estimated the proportion of internet users in each constituency, making use of figures generated by Blank, Graham and Calvino (2017). Appendix A1.1 contains further details of this estimation. As the estimations are based on census data (which is only updated once every decade), the indicators are the same for the 2015 and 2017 waves, which means that we cannot use this variable in our first difference models. As a second approach to measuring audience size, we also recorded the number of followers the candidate had during the campaign window. This number was observed each time the candidate sent a tweet (if the candidate sent multiple tweets, the final observed follower count was used). The internet usage statistics and follower counts allow us to address hypothesis 2.

## H3: Party Size

To address H3, we recorded the average vote share of each party in the electoral constituencies in which they competed. We use average share per constituency rather than total national vote share because regional parties such as the Scottish National Party performed very well where they were competing, but achieved only a fractional vote share overall. This variable is also used as a control variable in our first difference models, which allows us to distinguish between the impact of the individual candidate and the party at large.

---

[5] See: https://democracyclub.org.uk

[6] The validation was based on a random sample of the data (116 observations). For each observation in the sample, one of the authors made use of both the Twitter search function and Google to determine if the candidate in question had a Twitter username, and if so to look at if it had been correctly recorded in the dataset.

[7] In 2015, we collected the data using a commercial partner, Data Sift, which ensured that we did not encounter any rate limit issues. In 2017, we had elevated access to the Twitter Streaming API (Application Programming Interface) and again were able to collect the data without any rate limit issues. For both elections, we looked up the numeric ids corresponding to candidates' screennames using the Twitter RESTful API and used these to track their activity as incumbent MPs often change their Twitter usernames during the campaign period (during which they are not allowed to refer to themselves as MPs).

[8] In 2015 this ran from April 13th to May 6th, whilst in 2017 this ran from May 16th to June 7th.

### H4: Network Heterogeneity

To address H4, we recorded the number of times the candidate was 'mentioned' on Twitter (i.e. the number of times a tweet was sent containing the politician's username), and then calculated the proportion of these 'mentioners' who also mentioned a candidate in another party, which provides an estimate of the political heterogeneity of the candidate's audience (following a similar logic to Bright 2018).

### H5: Interactivity

Our interactivity hypothesis was addressed in two ways. First, we divided candidate tweeting activity into three major types: replies, retweets and 'original tweets'. We assume that the amount of replies is indicative of the amount of interaction a politician engages in. However, it is also possible that the original tweets have significant patterns of interactivity within them. Hence, as a second step, we fit an unsupervised topic models on all original tweets using non-negative matrix factorization (Gillis, 2014) for each party in each wave of observation. We selected the number of topics for each party using the method proposed by Greene, O'Callaghan & Cunningham (2014). One author of the study then coded each topic in each party as being related to either 'broadcasting' or 'interacting', based on the codebook developed by Graham et al. (2014). This process was repeated by a second author, and percent agreement was found to be 88%, with a Krippendorff's Alpha of 0.69. We then use our coded topic model to assign each tweet by each candidate into a topic, and hence calculate the amount of interacting and broadcasting communication the candidate engages in (a fuller explanation of the topic model process is provided in appendix A1.2).

### Control Variables

We also collected a number of control variables which we thought could be potential confounds of the Twitter effect. We included the party of the candidates themselves, whether they were currently an incumbent Member of Parliament [MP] or not in either the 2015 or 2017 elections, and the amount of money spent during the election campaign. Campaign spending data was made available by the UK Electoral Commission, a public body that monitors the conduct of elections[9]. Our spending data covers all money spent during the campaign period immediately before the election. However, at the time of writing the 2017 spending data did not include a handful of candidates who had yet to finish their returns. Hence models which include 2017 spending data do not include these observations.

### 4 Results

Selected descriptive statistics from the dataset are presented in Table 1 (full summary statistics for all variables are available in appendix A2). There are around 3,000 observations in both waves of observation, or around five per constituency (although we have included candidates from seven parties, not all of them field a candidate in each constituency). The lower overall number of observations in 2017 is largely driven by a considerable decrease in both Green Party and UKIP candidates. Twitter use is relatively widespread in both cohorts, though it drops from 76% in 2015 to 63% in 2017. This may be explained by the rushed nature of the 2017 election: many candidates were picked at the last minute, and hence had less time to organise social media profiles. Those candidates who did have a Twitter account made quite frequent use of the network in the campaign period, sending on average more than 150 tweets both waves of observation (though usage levels decreased slightly between 2017 and 2015). The majority of these were typically retweets, with interactive tweets being the least common type of communication, supporting work which has claimed that politicians mostly engage in 'broadcasting' activity.

---

[9] Available from: https://www.electoralcommission.org.uk/find-information-by-subject/elections-and-referendums/past-elections-and-referendums/uk-general-elections/candidate-election-spending

|  | 2015 | 2017 | 2015 and 2017 |
|---|---|---|---|
|  | *Mean (standard deviation) / Mean delta (SD delta)* | | |
| Tweets sent | 169.65 (243) | 152.94 (248) | -18.87 (183) |
| Interacting Tweets | 2.98 (10) | 7.88 (17) | 4.50 (12) |
| Broadcasting Tweets | 48.21 (75) | 43.63 (83) | -1.41 (68) |
| Retweets | 85.19 (155) | 84.92 (168) | -4.81 (124) |
| Replies | 37.52 (71) | 23.82 (58) | -15.51 (53) |
| Followers | 4,005.26 (26,201) | 8,261.67 (39,366) | 7,835 (43,689) |
|  |  |  |  |
| Number of candidates | 3,172 | 2,826 | 1,062 |
| Candidates with Twitter account | 2,398 (76%) | 1,786 (63%) | 822 (77%) |

*Table 1: **Descriptive Statistics.** Averages include only those candidates with a Twitter account*

Having a Twitter account was strongly and statistically significantly correlated with campaign spending and incumbency in both waves of observation; these factors were also correlated with the level of usage of these accounts and the amount of followers they have (though the effect sizes were relatively small in the case of usage levels). This offers support to the 'normalisation' thesis: established and well-resourced political forces seem to use the technology more and also attract more followers.

A final descriptively interest result concerns the geographic distribution of the activity of those mentioning the candidate, which we base on the portion of the tweets within our dataset came with latitude and longitude information attached. In both years an average of just over 30% of mentions of candidates came from their own constituency. This indicates that, while the majority of Twitter activity concerning candidates takes place outside their constituencies, a considerable proportion nevertheless takes place amongst people who could actually be eligible to vote for the candidate in question. Our data also allows us to perform a similar procedure for the candidate's own tweets: interestingly, we found that less than 50% of candidate tweets came from within their own constituency during this campaign period. However, we should note of course that only a small, non-representative subset of Twitter users enable the geolocation feature of the platform which allows us to address where the contribution comes from (Graham, Hale & Gaffney, 2014). Hence these estimates must be treated with caution.

Our analytical exercise consists of a set of linear models of the relationship between Twitter use and vote outcomes, contained in Table 2 and 3. All fitted models were analysed with standard goodness of fit diagnostic tests for OLS models[10]. These tests highlighted two main potential concerns with the fit: evidence of heteroscedasticity and a group of around 150 high influence observations (the exact amount varying depending on the model being fitted)[11]. These problems were not resolved by log transformation of variables, so two further steps were taken in response. First, coefficient estimates, measures of statistical significance and estimates of adjusted $R^2$ were all computed by bootstrapping (R=5,000). Second, robust linear regressions were also estimated for all models. The results of the robust regressions were the same as the original OLS estimates (with one exception which is noted below). Hence we have reported the original OLS estimates to facilitate interpretation.

---

[10] In particular, variance inflation factors were inspected for evidence of multicollinearity, Cook's distance was calculated for all observations in order to identify high influence data points, plots of variables versus fitted values were inspected to check the general assumption of a linear relationship between dependent and independent variables, and a Breusch-Pagan test for heteroscedasticity was conducted.

[11] The majority of these observations were parliamentary candidates who were campaigning in "safe" parliamentary constituencies, which are characterised by very high levels of concentration of voters for one party. In these safe seats, parliamentary candidates can score very high levels of vote share with little campaigning effort. Hence, these observations deviated considerably from the overall fit of the model. This impact was sometimes compounded when an MP retired and a new candidate took their place. In this situation, the new candidate is not treated as an "incumbent" by the model, but has many of the advantages of being an incumbent.

The models in Table 2 seek to tackle our fundamental question (hypothesis 1), of whether using Twitter increases vote share outcomes. Models 1.1 – 1.3 are cross-sectional, with the relationship between Twitter use and vote outcomes tested on both waves of our data, and hence in the table estimates are reported for both 2015 and 2017. Model 1.1 simply addresses the relationship between having a Twitter account and final vote share outcome. In both years Twitter use is positively associated with vote outcomes, though the overall adjusted $R^2$ of the models is small (0.04 in 2015 and 0.09 in 2017). Model 1.2 adds in our control variables: spending data, incumbency, and the party of the candidate. The party variable is not reported to save room, but full models can be found in appendix A3.1. We find the strength of the Twitter 'effect' decreases when compared to Model 1.1, but the variable itself remains significant (whilst adjusted $R^2$ increases considerably). In both models the effect of having a Twitter account is comparable with the effect of increasing spending (a little more in 2015, a little less in 2017). The absolute magnitude of the effect is modest: those with a Twitter account were between 1 and 2 percentage points better off in terms of vote share.

Model 1.3 is restricted to the subset of observations for candidates who did have a Twitter account, with the main independent variable being the amount of times they sent messages from this account during the observation window (log transformed). These messages include all original tweets, replies and retweets. The term is again positive and statistically significant, and remains modest but roughly comparable with spending: increasing the number of tweets sent by a factor of 10 increases vote share by around 1 percentage point. Model 1.4 pools all observations which had a Twitter account, and includes a dummy variable for whether the wave was 2017 or 2015 as well as an interaction term between the wave and the number of tweets sent. The Twitter term remains statistically significant, however the interaction term is not, indicating that the effect of Twitter in 2015 and 2017 is roughly comparable.

Model 1.5 offers our strongest test of the impact of Twitter on vote share, using a first difference model on the politicians who competed in both 2015 and 2017 and who had a Twitter account. All variables are now measured as differences between the 2015 and 2017 waves. The party term is the average vote share of the party overall, per constituency, differenced between 2015 and 2017: hence the impact of changes in the overall performance of individual parties is controlled for. Even in this strong test, increasing Twitter use between the two waves made a difference for politicians: candidates who increased their Twitter activity by a factor of 10 in 2017 typically improved on their 2015 score by just over half a percentage point. Interestingly, the amount of money spent is no longer statistically significant under this framework. This provides strong evidence that Twitter use does genuinely make a difference (though again the size of the effect is modest).

The second part of our analysis tackles hypothesis 2 to hypothesis 5, and seeks to understand the conditions whereby politicians gain more benefit from Twitter use. Table 3 contains linear models addressing different aspects of these hypotheses, following the same logic as Table 2: each model is tested on 2015 and 2017 data, and then as a first difference model. Each model is a duplication of model 1.3 above, with an added interaction term between the number of tweets sent and a new variable of interest. Only the coefficient for the interaction term is reported to save space (full models can be found in appendix A3.2). The number of observations fluctuates slightly in each model due to small amounts of missing data. A summary of the reasons for these fluctuations can be found in appendix A3.3.

| | Dependent variable: vote share | | | | | |
| | Model 1.1 | | Model 1.2 | | Model 1.3 | |
| | 2015 | 2017 | 2015 | 2017 | 2015 | 2017 |
|---|---|---|---|---|---|---|
| Uses Twitter | 8.68 *** | 13.43 *** | 1.82 *** | 1.06 ** | | |
| #Tweets ($\log_{10}$) | | | | | 0.96 *** | 0.85 *** |
| Spending ($\log_{10}$) | | | 0.95 *** | 2.49 *** | 0.96 *** | 3.68 *** |
| Incumbent MP | | | 23.35 *** | 25.37 *** | 22.62 *** | 24.17 *** |
| Party | | | Included | Included | Included | Included |
| Observations | 3,172 | 2,826 | 3,172 | 2,807 | 2,398 | 1,776 |
| Adjusted $R^2$ | 0.04 | 0.09 | 0.79 | 0.88 | 0.78 | 0.87 |

| | Model 1.4 | Model 1.5 Dependent variable: Δ vote share | |
|---|---|---|---|
| #Tweets ($\log_{10}$) | 0.74 ** | Δ Tweets (log10) | 0.57 * |
| Spending ($\log_{10}$) | 1.62 *** | Δ Spending (log10) | 0.03 |
| Incumbent MP | 22.66 *** | Δ Incumbent | 1.89 *** |
| Party | Included | Δ Party Vote Share | 1.09 *** |
| 2017 Wave | 0.40 | | |
| 2017 Interaction | 0.54 | | |
| Observations | 4,174 | | 818 |
| Adjusted $R^2$ | 0.81 | | 0.60 |

$^*$ $p < 0.05$ ; $^{**}$ $p < 0.01$; $^{***}$ $p < 0.001$

*Table 2: **OLS regressions of the relationship between Twitter use and vote share outcomes.** Coefficients, significance levels and R2 calculated using bootstrapping (percentile method, 5,000 repetitions). A categorical party variable is included but not reported in models 1.2-1.4 (full models can be found in the appendix)*

Hypotheses 2 concerns the effect of audience size. We address this in models 2.1 and 2.2. Model 2.1 incorporates an interaction term between our measure of internet use in a constituency, which we expect to broadly correlate with Twitter use. The term is statistically insignificant in both models. We cannot run a first difference model for internet use, as our measure is based on census data and is hence is identical for 2015 and 2017. Model 2.2 makes use of an interaction between the number of followers a politician has on Twitter and their tweeting activity. The term is again insignificant in both waves, and also in the difference model. Hence overall there is no supportive evidence for hypothesis 2.

Hypothesis 3 addressed the normalisation versus equalisation thesis. This is tackled in model 2.3, which incorporates an interaction term for the size of the politician's party, measured as the average vote share achieved by candidates in that party (note that the dummy variables for the party itself are of course dropped). There is strong evidence that party size makes a difference: the term is significant in 2015 and in the difference model. However it is positive, meaning that we reject the notion that small parties benefit more. Indeed, larger parties seem to do better. This is evidence for 'normalisation' theory. Of course, we should highlight that there are only seven parties in the study, and hence the variation we have on this variable is limited. Multi-country research would be needed to fully confirm this finding.

Hypothesis 4 relates to the 'echo chamber' effect, stating that politicians with more diverse audiences should gain more from tweeting. We tackle this in model 2.4, which interacts the percentage of 'multi-party' mentioners that a politician has with the volume of tweets they produce.

The term is strongly significant and negatively signed in the 2015 wave, but is not significant in the 2017 wave or in the difference model. Furthermore, the direction of effect is reversed in the difference model when compared with the cross sectional models. Therefore, there is overall little evidence to support hypothesis 4.

Hypothesis 5, finally, concerns the extent to which politicians engage with Twitter in an interactive manner. Model 2.5 looks at this by breaking down Twitter activity into the three major types of tweet: replies (where the user responds directly to a comment from another user), retweets (where the user rebroadcasts a message from another user) and what we have labelled here as 'original tweets' which include everything else. There is no evidence that replies have more of an impact than any other kind of tweet. There is a little evidence by contrast that original tweets make a difference, with the term significant in 2017.

Model 2.6 splits the original tweets produced by the politician into 'interacting' and 'broadcasting' categories using the results of our topic model, as outlined above. Here, surprisingly, there is strong evidence that broadcasting tweets make the most difference, with the term statistically significant in both 2015 and the first difference model. The result is reversed in 2017, though this finding is not significant in the robust regression model.

| | Dependent variable: vote share for 2015 and 2017 models, difference for FD models | | | | | | | | |
|---|---|---|---|---|---|---|---|---|---|
| | H2: Audience Size | | | | | | H3: Norm. vs. Equalisation | | |
| | Model 2.1 | | | Model 2.2 | | | Model 2.3 | | |
| Term (Δ for FD models) | 2015 | 2017 | FD | 2015 | 2017 | FD | 2015 | 2017 | FD |
| Internet Use * Tweets (log$_{10}$) | 0.26 | 0.37 | NA | | | | | | |
| Followers (log$_{10}$) * Tweets (log$_{10}$) | | | | 0.73 | 0.87 | -0.53 | | | |
| Avg. Party Share * Tweets (log$_{10}$) | | | | | | | 0.05** | 0.02 | 0.09* |
| Observations | 2,398 | 1,776 | NA | 2,274 | 1,569 | 723 | 2,398 | 1,776 | 818 |
| Adjusted R$^2$ | 0.78 | 0.87 | NA | 0.79 | 0.88 | 0.60 | 0.77 | 0.87 | 0.69 |

| | H4: Echo Chamber | | | | | | H5: Interactivity | | |
|---|---|---|---|---|---|---|---|---|---|
| | Model 2.4 | | | Model 2.5 | | | Model 2.6 | | |
| | 2015 | 2017 | FD | 2015 | 2017 | FD | 2015 | 2017 | FD |
| Perc. MP * Tweets (log$_{10}$) | -8.64*** | -1.76 | 0.83 | | | | | | |
| Original Tweets | | | | 0.69 | 1.26* | 0.72 | | | |
| Replies | | | | 0.43 | 0.47 | 0.62 | | | |
| Retweets | | | | 0.12 | -0.55 | -0.52 | | | |
| Interacting | | | | | | | 0.60 | 1.81** | -0.43 |
| Broadcasting | | | | | | | 1.15** | 0.29 | 1.33** |
| Observations | 2,319 | 1,712 | 788 | 2,398 | 1,776 | 818 | 2,199 | 1,523 | 692 |
| Adjusted R$^2$ | 0.79 | 0.87 | 0.68 | 0.78 | 0.87 | 0.69 | 0.77 | 0.87 | 0.59 |

* $p < 0.05$ ; ** $p < 0.01$ ; *** $p < 0.001$

*Table 3: **Further OLS Models** Terms for incumbency and spending are included in all models but not reported. A party term is also included in all models apart from 2.3. Grey shading indicates a result which was not significant in the robust regressions.*

## 5 Discussion

This study has provided, to our knowledge, the strongest empirical test yet of the link between political campaigning on Twitter and vote share outcomes. The results show considerable evidence that there is a connection between political Twitter activity and votes: a variety of different model specifications were tested, with relevant control variables for spending and party membership, on two waves of panel data, pooled time series and first difference models. In all of these a positive, statistically significant correlation was found. However, the results were also unanimous in suggesting that the impact of Twitter use is small in absolute terms. In our cross sectional models, candidates that had a Twitter account typically had vote shares around 1-2 percentage points higher than those who did not (see model 1.2). Amongst those that did have an account, a 10-fold increase in the volume of tweets sent would be needed to generate a 1 percentage point increase in vote share (model 1.3). In our first difference model the effect was event smaller: a 10-fold increase in tweets sent would generate just over half a percentage point of vote share increase.

The relative size of the effect should be taken into account. Twitter use was comparable with the effect of campaign spending in our cross sectional models, and outstrips it in our differenced regressions. It is also worth highlighting that modest increase in vote share may also be significant in a close race, of which there were many in our data: around 14% of the electoral competitions which form the basis of our study were won by a margin of less than 5 percentage points, and 4% of them were won by a margin of less than 1 percentage point. However, considering the small size of the effect, and also that the effort of producing new tweets is essentially constant regardless of how many are produced, the results suggest that the best strategy for a political candidate is a limited amount of engagement with the platform (say, between 100 and 1,000 tweets sent during the campaign period).

We also addressed the circumstances under which political twitter use was particularly effective. We found little evidence that the size of a candidate's Twitter audience made a difference, or that the heterogeneity of that audience had an impact. We also found little evidence that making use of interactive conversation strategies was impactful: indeed there was evidence that engaging with Twitter in 'broadcast' mode was more helpful. This challenges existing work which has suggested that interactive Tweeting patterns generate a vote dividend, and indeed the large body of literature which has criticised politicians for not engaging in social media in a more interactive fashion. There was also good evidence that larger parties benefitted more from using the technology, contradicting the 'equalisation' thesis whilst supporting the 'normalisation' thesis.

One way of summarising these interaction effects is that they lend weight to the idea that campaigning on Twitter is more likely to produce *indirect* benefits rather than *direct* ones. That is to say, use of Twitter may have an impact because it stimulates coverage in media outlets, or because it energises opinion leaders, but not because it creates direct connections between politicians and voters. This would explain why the size and nature of the politician's audience does not make a difference, nor do their attempts to interact with these audiences. Rather, broadcast tweets which might lend themselves to further media coverage are positioned to generate an impact. This would also explain why larger parties benefit more: following scholars in the normalisation tradition, we speculate that the existing resource advantages enjoyed by established parties also translate into this new technology. This might include more access to professional campaign staff who can advise on the use of new media technology, or a greater likelihood that their Twitter activities are covered by journalists who perceive them as more important.

It is worth concluding by highlighting weaknesses in the study, and thus pointing the way for future research. We would highlight two of these in particular. First, we address only one social media platform in this paper. We do not know the extent to which the use of Twitter correlates with use of other types of social media (such as Facebook and Snapchat), hence we are unable to say to what extent it is Twitter itself which makes the difference, as compared to other platforms. Future work that studied campaign effort on multiple platforms would be highly valuable. Second, we treat observations in our dataset as if they are the result of separate, independent electoral campaigns. This is, of course, not the case. Candidates face each other in constituencies: two brilliant campaigns might cancel each other out, and hence produce no result in our data. There are also co-ordinated national campaigns that take place containing social media elements. Though our party variables and first difference regressions address these problems to an extent, they cannot resolve them fully. Only a field experiment could truly address these deficiencies. Absent this, however, we believe that this study is the strongest treatment yet of the effect of active political campaigning on social media.

## Appendix

This appendix is divided into three parts: in section A1 we give full descriptive statistics for our dataset; in section A2 we provide further details on how some of our variables were operationalised; and in section A3 we report in full models which are only described in abridged form in the body text of the paper.

## A1 Variable Operationalisation

In this section of the appendix, we provide more details on how our constituency level internet use indicator is operationalised, and the approach we take to constructing and coding our topic models.

### A1.1 Constituency Level Internet Use Indicators

Our constituency level internet use indicators are based on the data released by Blank, Graham & Calvino (2017). This paper used small area estimation to provide a measure of internet use at the "output area" level, which is a small geographical area designed specifically for use with the census which typically contains around 120 households. There are 227,759 of these output areas in England, Scotland and Wales. The census provides lookup tables which map these output areas to parliamentary constituencies. We use these lookup tables to provide a population weighted average of internet usage by constituency on the basis of the output area measures. In cases where output areas overlapped with two or more constituencies, we assigned these areas to the constituency with which they had the largest overlap.

### A1.2 Topic Models

In order to characterise the type of communication activity which political candidates engage in during the campaign period on Twitter, we make use of a series of topic models, a technique which is increasingly used in communication research (Maier et al., 2018). The approach allows us to extract a discrete number of general topics from the textual data within candidate tweets. The advantage of unsupervised topic models over methods such as content analysis and supervised machine learning is that they do not require a large amount of up front data, and therefore they make tractable the process of characterising the communication style of thousands of different political candidates. We will briefly describe the process we followed to produce our topic models here.

The input into a topic model is a 'document-term matrix', which is simply a matrix whereby each row is an individual document (an individual tweet in our case) and each column is a word which appears in the entire corpus of tweets. The entry $d\text{-}t_{ij}$ specifies the amount of times the term $t_j$ appears in the document $d_i$. We only made use of 'original' tweets in our corpus (i.e. we do not include replies and retweets). We pre-process our corpus of tweets to remove common 'stopwords' (frequently occurring words such as 'and', 'the', 'it' etc. which we assume have little value in classification terms). We also 'lemmatize' all remaining words in the corpus, returning each word to its original base or lemma (such that, for example the words 'angry' and 'angrier' would be reduced to the same term, angry). We then limit our model to the 1,000 most frequently occurring words in the corpus, on the basis that extremely infrequent words are unlikely to be of use in distinguishing topics. Finally, we convert these term frequency scores using tf-idf weighting, which applies a stronger weight to terms which are less common across the corpus of the documents as a whole. From this document-term matrix, two further matrices are estimated: a term-topic matrix (which specifies the likelihood of individual terms appearing in a given topic) and a document-topic matrix (which specifies the likelihood of a given document, or tweet, belonging to a particular topic). These matrices were estimated using the non-negative matrix factorization [NMF] technique (see e.g. Gillis, 2014). We used NMF rather than the slightly more popular latent dirichlet allocation because NMF allows us to work with fractional tf-idf scores.

A key consideration in topic models is choosing the appropriate number of topics, $k$, which must be specified by the researcher (Maier et al., 2018). The ideal value of $k$ is one that allows the full variation of different communication styles to be captured without creating arbitrary divisions between groups of documents that are in practice quite similar. We chose to fit an individual topic model for each party in each year, on the basis that different parties might choose different communication strategies for their campaigns. To select the appropriate number of topics for each campaign year, we made use of the stability analysis technique proposed by Greene, O'Callaghan & Cunningham (2014), which is adapted to the particular case of NMF and involves analysing the extent to which the topic-term matrix is robust to random perturbations of the input data for different values of $k$. For all parties in each of the two waves of our data, we tested all values of $k$ between 5 and 15 against 10 randomly drawn samples of 80% of the data. We then picked the value of $k$ with the highest level of stability for each party (in the end all values of $k$ were in the range 5-7).

Once we had created each party-year level topic model, we then labelled each topic in terms of its style of communication: where it could be conceived of as largely one-way communication (*broadcasting*) or whether it attempted to engage in some form of two-way communication (*interacting*). Our definitions of these two communication types come from the codebook proposed in Graham et al. (2014, pp.697-698), where they label a number of different types of communication as falling into one of these two overarching categories. The definitions we use are set out in table A1.

| Broadcasting Behaviours | Definition |
| --- | --- |
| Updating | Posting updates on recent candidate activity, e.g. attendance at events or doorstep campaigning |
| Promoting | Tweets specifically promoting the skills/ability of the candidate or party |
| Critiquing | Tweets criticising other parties or candidates |
| Information disseminating | Dissemination of news reports or informational links |
| Own / party stance | Tweets where the candidate takes a position on a specific policy area |
|  |  |
| Interacting beahviours | |
| Debating | Tweets where the candidate directly debates with opposition candidates or members of the general public |
| Acknowledging | Tweets where the candidate thanks people or acknowledges support |
| Organizing / mobilizing | Direct efforts to organise offline or online activity |
| Advice giving / helping | Candidate efforts to help people in individual constituencies |
| Consulting | Requesting public input on a given issue |

*Table A1. Topic codebook. Adapted from Graham et al., 2014, pp.703-707.*

The coding process itself was based on both the term-topic matrix and the document-topic matrix. For each topic, we extracted the 10 most probable terms and the 5 most probable tweets. One of the authors of the study then used the content of these terms and tweets to make a coding decision. A second author independently performed the process, producing a Krippendorrf's alpha of 0.69 (percent agreement of 88%). Having labelled all topics as either *broadcasting* or *interacting,* we were then able to label all tweets in the corpus as either *broadcasting* or *interacting* as well, on the basis of selecting the most probable topic for each tweet.

## A2 Descriptive Statistics

| *Categorical variables* | 2015 freq | % | 2017 freq | % | 2015 - 2017 Panel | freq | % |
|---|---|---|---|---|---|---|---|
| Total observations | 3,172 | 100 | 2,826 | 100 | | 1062 | 100 |
| | | | | | | | |
| Incumbent MP | 540 | 17% | 597 | 21% | Inc. 2015, Chal. 2017 | 30 | 3% |
| Challenger MP | 2,632 | 83% | 2,229 | 79% | No Change | 875 | 81% |
| | | | | | Chal. 2015, Inc. 2017 | 157 | 16% |
| | | | | | | | |
| Conservative Party | 631 | 20% | 631 | 22% | Conservative Party | 359 | 35% |
| Green Party | 567 | 18% | 459 | 16% | Green Party | 105 | 10% |
| Labour Party | 631 | 20% | 631 | 22% | Labour Party | 285 | 31% |
| Liberal Democrats | 631 | 20% | 629 | 22% | Liberal Democrats | 191 | 17% |
| Plaid Cymru | 40 | 1% | 40 | 1% | Plaid Cymru | 3 | 0% |
| Scottish National Party | 59 | 2% | 59 | 2% | Scottish National Party | 46 | 4% |
| UK Independence Party | 613 | 19% | 377 | 13% | UK Independence Party | 73 | 4% |
| | | | | | | | |
| Had Twitter | 2,398 | 76% | 1,786 | 63% | Had Twitter | 822 | 77% |
| Did not have Twitter | 774 | 24% | 1,040 | 37% | Did not have Twitter | 240 | 23% |
| | | | | | | | |
| *Numeric variables* | mean | sd | mean | sd | | mean | sd |
| Vote Share | 19.73 | 17.79 | 22.17 | 22.20 | Share 2017 - 2015 | 2.79 | 8.31 |
| Internet Use in Cons. | 0.78 | 0.06 | 0.78 | 0.06 | | | |
| | | | | | | | |
| *Numeric variables (those with Twitter account only)* | | | | | | | |
| Tweets | 169.65 | 242.72 | 152.94 | 248.53 | Tweets 2017 - 2015 | -18.87 | 183.33 |
| Spending | 3877.99 | 4586.52 | 4709.97 | 4676.83 | Spending 2017 - 2015 | 1061.48 | 3818.90 |
| Followers | 4005.26 | 26201.59 | 8261.67 | 39366.16 | Followers 2017 - 2015 | 7834.59 | 43688.96 |
| Party Av. Share | 19.73 | 13.63 | 22.17 | 18.68 | Party Av. 2017 - 2015 | 2.97 | 6.51 |
| % Multi Party Mentioners | 0.71 | 0.17 | 0.84 | 0.11 | %MP 2017 - 2015 | 0.16 | 0.17 |
| Original Tweets | 46.95 | 76.20 | 44.19 | 84.40 | Orig. 2017 - 2015 | 1.45 | 71.47 |
| Replies | 37.52 | 70.91 | 23.82 | 58.05 | Replies 2017 - 2015 | -15.51 | 52.61 |
| Retweets | 85.19 | 154.94 | 84.92 | 167.86 | Retweets 2017 - 2015 | -4.81 | 124.24 |
| Broadcast Tweets | 48.21 | 74.76 | 43.63 | 83.07 | Broadcast 2017 - 2015 | -1.41 | 68.21 |
| Interacting Tweets | 2.98 | 10.33 | 7.88 | 17.40 | Interacting 2017 - 2015 | 4.50 | 11.98 |

*Table A2: Descriptive Statistics*

## A3 Models

This section contains full models from tables 2 and 3.

*A3.1 Full Models from Table 2*

| 2015 Models | Model 1.1 | | | Model 1.2 | | | Model 1.3 | | |
|---|---|---|---|---|---|---|---|---|---|
| | Coef | 95% CI | | Coef | 95% CI | | Coef | 95% CI | |
| Uses Twitter | 8.68 | 7.39 | 9.94 | 1.82 | 1.23 | 2.42 | | | |
| #Tweets (log10) | | | | | | | 0.96 | 0.52 | 1.41 |
| Spending (log10) | | | | 0.95 | 0.72 | 1.18 | 0.96 | 0.68 | 1.26 |
| Incumbent MP | | | | 23.35 | 22.28 | 24.43 | 22.62 | 21.42 | 23.78 |
| Green Party | | | | -21.17 | -22.29 | -20.04 | -22.00 | -23.23 | -20.76 |
| Labour Party | | | | -2.61 | -3.88 | -1.31 | -2.74 | -4.12 | -1.44 |
| Liberal Democrats | | | | -19.63 | -20.75 | -18.52 | -19.74 | -20.97 | -18.53 |
| Plaid Cymru - The Party of Wales | | | | -14.88 | -17.44 | -12.05 | -15.05 | -18.04 | -11.72 |
| Scottish National Party (SNP) | | | | 20.28 | 17.69 | 22.79 | 19.54 | 17.01 | 22.02 |
| UK Independence Party (UKIP) | | | | -11.98 | -13.17 | -10.78 | -12.67 | -14.01 | -11.34 |
| Adjusted R2 | 0.04 | | | 0.79 | | | 0.78 | | |
| N | 3,172 | | | 3,172 | | | 2,398 | | |

| 2017 Models | Model 1.1 | | | Model 1.2 | | | Model 1.3 | | |
|---|---|---|---|---|---|---|---|---|---|
| | Coef | 95% CI | | Coef | 95% CI | | Coef | 95% CI | |
| Uses Twitter | 13.44 | 11.94 | 14.97 | 1.06 | 0.44 | 1.66 | | | |
| #Tweets (log10) | | | | | | | 0.85 | 0.40 | 1.32 |
| Spending (log10) | | | | 2.49 | 2.21 | 2.79 | 3.68 | 3.15 | 4.29 |
| Incumbent MP | | | | 25.37 | 24.29 | 26.49 | 24.18 | 22.87 | 25.45 |
| Green Party | | | | -23.30 | -24.36 | -22.23 | -22.73 | -24.16 | -21.26 |
| Labour Party | | | | 3.46 | 2.35 | 4.53 | 3.60 | 2.35 | 4.86 |
| Liberal Democrats | | | | -21.44 | -22.60 | -20.28 | -21.03 | -22.47 | -19.56 |
| Plaid Cymru - The Party of Wales | | | | -19.73 | -21.74 | -17.54 | -18.91 | -21.46 | -16.22 |
| Scottish National Party (SNP) | | | | -16.08 | -18.07 | -13.93 | -16.09 | -18.47 | -13.31 |
| UK Independence Party (UKIP) | | | | -22.40 | -23.49 | -21.31 | -22.44 | -24.05 | -20.83 |
| Adjusted R2 | 0.09 | | | 0.88 | | | 0.87 | | |
| N | 2,826 | | | 2,807 | | | 1,776 | | |

| Pooled and First Difference Model | Model 1.4 | | | | Model 1.5 | | |
|---|---|---|---|---|---|---|---|
| | Coef | 95% CI | | | Coef | 95% CI | |
| #Tweets (log10) | 0.74 | 0.27 | 1.20 | Δ Tweets (log10) | 0.57 | 0.11 | 1.04 |
| Spending (log10) | 1.62 | 1.35 | 1.89 | Δ Spending (log10) | 1.89 | 0.91 | 2.85 |
| Incumbent MP | 22.66 | 21.73 | 23.57 | Δ Incumbent | 0.03 | -0.25 | 0.31 |
| Green Party | -23.38 | -24.35 | -22.40 | Δ Party Vote Share | 1.09 | 1.04 | 1.13 |
| Labour Party | 0.01 | -1.00 | 1.03 | | | | |
| Liberal Democrats | -20.97 | -21.95 | -20.00 | | | | |
| Plaid Cymru - The Party of Wales | -17.41 | -19.53 | -15.25 | | | | |
| Scottish National Party (SNP) | 4.01 | 0.38 | 7.64 | | | | |
| UK Independence Party (UKIP) | -16.23 | -17.39 | -15.07 | | | | |
| Year | 0.39 | -0.87 | 1.65 | | | | |
| Year * #Tweets (log10) | 0.55 | -0.13 | 1.23 | | | | |
| Adjusted R2 | 0.79 | | | | 0.68 | 0.64 | 0.73 |
| N | 4,174 | | | | 818 | | |

*Table A3: Full models from table 2. Conservative Party is the reference level for the party variable.*

*A3.2 Full Models from Table 3*

| **2015 Models** | **Model 2.1** | | | **Model 2.2** | | | **Model 2.3** | | |
|---|---|---|---|---|---|---|---|---|---|
| | Coef | 95% CI | | Coef | 95% CI | | Coef | 95% CI | |
| #Tweets (log10) | 0.89 | 0.44 | 1.35 | -3.16 | -5.13 | -1.15 | 0.06 | -0.51 | 0.60 |
| Spending (log10) | 0.98 | 0.69 | 1.27 | 0.75 | 0.45 | 1.05 | 0.96 | 0.67 | 1.24 |
| Incumbent MP | 22.62 | 21.47 | 23.77 | 17.63 | 16.24 | 18.99 | 21.65 | 20.56 | 22.79 |
| Green Party | -22.04 | -23.29 | -20.83 | -21.16 | -22.38 | -19.95 | | | |
| Labour Party | -2.71 | -4.07 | -1.34 | -3.29 | -4.60 | -1.92 | | | |
| Liberal Democrats | -19.82 | -21.03 | -18.60 | -19.25 | -20.48 | -18.07 | | | |
| Plaid Cymru - The Party of Wales | -14.47 | -17.40 | -11.13 | -15.28 | -18.19 | -11.94 | | | |
| Scottish National Party (SNP) | 19.88 | 17.37 | 22.36 | 17.88 | 15.23 | 20.39 | | | |
| UK Independence Party (UKIP) | -12.68 | -14.06 | -11.35 | -12.83 | -14.18 | -11.52 | | | |
| Internet Use in Cons. | 0.06 | -0.80 | 0.97 | | | | | | |
| Internet Use in Cons. * #Tweets (log10) | 0.25 | -0.23 | 0.73 | | | | | | |
| Followers (log10) | | | | 3.44 | 1.89 | 5.03 | | | |
| Followers (log10) * #Tweets (log10) | | | | 0.73 | -0.04 | 1.45 | | | |
| Average Party Share | | | | | | | 0.62 | 0.54 | 0.70 |
| Average Party Share * #Tweets (log10) | | | | | | | 0.05 | 0.01 | 0.09 |
| Adjusted R2 | 0.78 | | | 0.79 | | | 0.77 | | |
| N | 2,398 | | | 2,274 | | | 2,398 | | |

| **2015 Models** | **Model 2.4** | | | **Model 2.5** | | | **Model 2.6** | | |
|---|---|---|---|---|---|---|---|---|---|
| | Coef | 95% CI | | Coef | 95% CI | | Coef | 95% CI | |
| #Tweets (log10) | 6.69 | 4.57 | 8.73 | | | | | | |
| Spending (log10) | 0.89 | 0.60 | 1.18 | 0.96 | 0.68 | 1.25 | 0.96 | 0.64 | 1.28 |
| Incumbent MP | 21.74 | 20.59 | 22.88 | 22.61 | 21.44 | 23.74 | 22.27 | 21.10 | 23.44 |
| Green Party | -21.66 | -22.89 | -20.47 | -22.02 | -23.31 | -20.79 | -22.29 | -23.81 | -20.77 |
| Labour Party | -2.98 | -4.31 | -1.65 | -2.75 | -4.10 | -1.38 | -2.68 | -4.25 | -1.09 |
| Liberal Democrats | -18.60 | -19.84 | -17.41 | -19.76 | -21.03 | -18.55 | -19.72 | -21.09 | -18.37 |
| Plaid Cymru - The Party of Wales | -12.41 | -15.29 | -9.19 | -15.12 | -18.01 | -11.78 | -15.18 | -18.20 | -11.77 |
| Scottish National Party (SNP) | 22.19 | 19.61 | 24.70 | 19.47 | 16.88 | 22.03 | 19.37 | 16.50 | 22.23 |
| UK Independence Party (UKIP) | -10.53 | -11.94 | -9.17 | -12.68 | -14.05 | -11.34 | -12.84 | -14.29 | -11.38 |
| % Multi-Party Mentioners | 0.54 | -3.56 | 4.35 | | | | | | |
| % MP. Mens. * #Tweets (log10) | -8.64 | -11.19 | -5.98 | | | | | | |
| Replies (log10) | | | | 0.43 | -0.41 | 1.29 | | | |
| Retweets (log10) | | | | 0.12 | -0.61 | 0.84 | | | |
| Original tweets (log10) | | | | 0.69 | -0.10 | 1.46 | | | |
| Interacting tweets (log10) | | | | | | | 0.60 | -0.67 | 1.87 |
| Broadcasting tweets (log10) | | | | | | | 1.15 | 0.33 | 1.96 |
| Adjusted R2 | 0.79 | | | 0.78 | | | 0.77 | | |
| N | 2,319 | | | 2,398 | | | 2,199 | | |

*Table A4: Full models from table 3, 2015 wave. Conservative Party is the reference level for the party variable.*

| 2017 Models | Model 2.1 | | | Model 2.2 | | | Model 2.3 | | |
|---|---|---|---|---|---|---|---|---|---|
| | Coef | 95% CI | | Coef | 95% CI | | Coef | 95% CI | |
| #Tweets (log10) | 0.88 | 0.42 | 1.35 | -3.49 | -6.37 | -0.67 | 0.61 | 0.11 | 1.11 |
| Spending (log10) | 3.70 | 3.15 | 4.29 | 3.30 | 2.78 | 3.90 | 3.36 | 2.84 | 3.94 |
| Incumbent MP | 24.22 | 22.93 | 25.49 | 18.55 | 16.80 | 20.26 | 22.47 | 21.19 | 23.73 |
| Green Party | -22.69 | -24.16 | -21.28 | -22.54 | -24.03 | -21.04 | | | |
| Labour Party | 3.62 | 2.31 | 4.93 | 4.03 | 2.72 | 5.31 | | | |
| Liberal Democrats | -21.02 | -22.45 | -19.57 | -20.56 | -22.01 | -19.08 | | | |
| Plaid Cymru - The Party of Wales | -18.95 | -21.55 | -16.27 | -18.31 | -21.06 | -15.40 | | | |
| Scottish National Party (SNP) | -16.06 | -18.52 | -13.43 | -16.27 | -18.44 | -14.00 | | | |
| UK Independence Party (UKIP) | -22.37 | -24.02 | -20.71 | -23.09 | -24.76 | -21.35 | | | |
| Internet Use in Cons. | -0.70 | -1.53 | 0.08 | | | | | | |
| Internet Use in Cons. * #Tweets (log10) | 0.37 | -0.06 | 0.81 | | | | | | |
| Followers (log10) | | | | 3.27 | 1.21 | 5.29 | | | |
| Followers (log10) * #Tweets (log10) | | | | 0.87 | -0.05 | 1.82 | | | |
| Average Party Share | | | | | | | 0.62 | 0.57 | 0.67 |
| Average Party Share * #Tweets (log10) | | | | | | | 0.02 | -0.01 | 0.04 |
| Adjusted R2 | 0.87 | | | 0.88 | | | 0.85 | | |
| N | 1,776 | | | 1,569 | | | 1,776 | | |

| 2017 Models | Model 2.4 | | | Model 2.5 | | | Model 2.6 | | |
|---|---|---|---|---|---|---|---|---|---|
| | Coef | 95% CI | | Coef | 95% CI | | Coef | 95% CI | |
| #Tweets (log10) | 2.23 | -0.88 | 5.49 | | | | | | |
| Spending (log10) | 3.63 | 3.10 | 4.24 | 3.63 | 3.11 | 4.25 | 3.73 | 3.15 | 4.41 |
| Incumbent MP | 23.79 | 22.50 | 25.09 | 24.18 | 22.89 | 25.48 | 23.70 | 22.30 | 25.06 |
| Green Party | -22.62 | -24.06 | -21.18 | -22.86 | -24.29 | -21.39 | -24.25 | -26.31 | -22.20 |
| Labour Party | 3.86 | 2.58 | 5.14 | 3.46 | 2.19 | 4.75 | 3.81 | 2.42 | 5.17 |
| Liberal Democrats | -20.65 | -22.10 | -19.16 | -21.16 | -22.58 | -19.68 | -20.74 | -22.36 | -19.14 |
| Plaid Cymru - The Party of Wales | -18.63 | -21.16 | -16.01 | -18.81 | -21.36 | -16.02 | -19.54 | -22.11 | -16.92 |
| Scottish National Party (SNP) | -15.25 | -17.78 | -12.58 | -16.26 | -18.65 | -13.66 | -17.68 | -20.07 | -15.13 |
| UK Independence Party (UKIP) | -22.12 | -23.71 | -20.50 | -22.62 | -24.19 | -21.00 | -22.83 | -24.67 | -21.00 |
| % Multi-Party Mentioners | -3.60 | -8.67 | 1.50 | | | | | | |
| % MP. Mens. * #Tweets (log10) | -1.76 | -5.48 | 1.79 | | | | | | |
| Replies (log10) | | | | 0.59 | -0.36 | 1.52 | | | |
| Retweets (log10) | | | | -0.52 | -1.36 | 0.34 | | | |
| Original tweets (log10) | | | | 1.26 | 0.23 | 2.27 | | | |
| Interacting tweets (log10) | | | | | | | 1.81 | 0.42 | 3.20 |
| Broadcasting tweets (log10) | | | | | | | 0.31 | -0.91 | 1.51 |
| Adjusted R2 | 0.87 | | | 0.87 | | | 0.87 | | |
| N | 1,712 | | | 1,776 | | | 1,523 | | |

*Table A5: Full models from table 3, 2017 wave. Conservative Party is the reference level for the party variable.*

| FD Models | Model 2.1 | | Model 2.2 | | | Model 2.3 | | |
|---|---|---|---|---|---|---|---|---|
| *All variables are 2015 - 2017 deltas* | Coef | 95% CI | Coef | 95% CI | | Coef | 95% CI | |
| #Tweets (log10) | | | 0.98 | 0.12 | 1.87 | 0.33 | -0.14 | 0.80 |
| Incumbency | | | 1.73 | 0.45 | 2.94 | 1.85 | 0.86 | 2.83 |
| Spending (log10) | NA: Internet use variable was the same in 2015 and 2017 | | 0.08 | -0.23 | 0.40 | 0.03 | -0.25 | 0.31 |
| Average Party Share | | | 1.10 | 1.05 | 1.15 | 1.10 | 1.05 | 1.14 |
| Followers (log10) | | | 0.12 | -1.11 | 1.39 | | | |
| Followers (log10) * #Tweets (log10) | | | -0.60 | -2.19 | 0.81 | | | |
| Average Party Share * #Tweets (log10) | | | | | | 0.09 | 0.02 | 0.16 |
| Adjusted R2 | | | 0.68 | | | 0.69 | | |
| N | | | 723 | | | 818 | | |

| FD Models | Model 2.4 | | | Model 2.5 | | | Model 2.6 | | |
|---|---|---|---|---|---|---|---|---|---|
| *All variables are 2015 - 2017 deltas* | Coef | 95% CI | | Coef | 95% CI | | Coef | 95% CI | |
| #Tweets (log10) | 0.35 | -0.23 | 1.00 | | | | | | |
| Incumbency | 1.90 | 0.87 | 2.92 | 1.89 | 0.94 | 2.90 | 1.73 | 0.75 | 2.75 |
| Spending (log10) | 0.03 | -0.25 | 0.32 | 0.02 | -0.28 | 0.29 | 0.05 | -0.27 | 0.39 |
| Average Party Share | 1.09 | 1.04 | 1.13 | 1.09 | 1.04 | 1.14 | 1.10 | 1.05 | 1.15 |
| % Multi-Party Mentioners | 0.52 | -0.97 | 2.04 | | | | | | |
| % MP. Mens. * #Tweets (log10) | 0.83 | -1.30 | 2.67 | | | | | | |
| Replies (log10) | | | | -0.42 | -1.08 | 0.23 | | | |
| Retweets (log10) | | | | 0.72 | -0.03 | 1.50 | | | |
| Original tweets (log10) | | | | 0.62 | -0.09 | 1.34 | | | |
| Interacting tweets (log10) | | | | | | | -0.43 | -1.33 | 0.45 |
| Broadcasting tweets (log10) | | | | | | | 1.33 | 0.58 | 2.07 |
| Adjusted R2 | 0.68 | | | 0.69 | | | 0.68 | | |
| N | 788 | | | 818 | | | 692 | | |

*Table A4: Full models from table 3, first difference models only.*

### A3.3 Observations Per Model

Due to the way variables are calculated the number of observations changes slightly in the models in Table 3. Here we describe the reasons for these changes.

**Model 2.1:** contains all observations which had a Twitter account and which had complete spending data.

**Model 2.2:** Follower counts were observed when candidates sent a tweet. Hence the model contains all observations which had a Twitter account, complete spending data and sent at least one Tweet.

**Model 2.3:** contains all observations which had a Twitter account and which had complete spending data.

**Model 2.4:** contains all observations which had a Twitter account, complete spending data and who were mentioned at least once. The proportion of 'multi-party' mentioners could only be calculated if the candidate was mentioned at least once.

**Model 2.5:** contains all observations which had a Twitter account and which had complete spending data.

**Model 2.6:** contains all observations which had a Twitter account, complete spending data and sent at least one 'original' tweet (i.e. something that was not either a reply or a retweet).